\documentclass[twocolumn,preprintnumbers,amsmath,amssymb]{revtex4}

\usepackage{graphicx}
\usepackage{dcolumn}
\usepackage{bm}
\usepackage[latin1]{inputenc}
\usepackage{color}
\usepackage{wasysym}

\begin{document}

\title{Topology and manipulation of multiferroic hybrid domains in MnWO$_4$}

\author{D. Meier}
\author{N. Leo}
\author{M. Maringer}
\author{Th. Lottermoser}
\author{M. Fiebig}
\affiliation{%
Helmholtz-Institut f\"ur Strahlen- und Kernphysik, Universit\"at Bonn\\
Nussallee 14 - 16, 53115 Bonn, Germany
}%

\author{P. Becker}
\author{L. Bohat\'y}
\affiliation{
Institut f\"ur Kristallographie, Universität zu K\"oln\\
Z\"ulpicher Strasse 49b, 50937 Köln, Germany
}%
\date{\today}

\begin{abstract}
An investigation of the spatially resolved distribution of domains in the multiferroic phase of
MnWO$_4$ reveals that characteristic features of magnetic and ferroelectric domains are
inseparably entangled. Consequently, the concept of {\it multiferroic hybrid domains} is
introduced for compounds in which ferroelectricity is induced by magnetic order. The
three-dimensional structure of the domains is resolved. Annealing cycles reveal a topological
memory effect that goes beyond previously reported memory effects and allows one to reconstruct
the entire multiferroic multidomain structure subsequent to quenching it.
\end{abstract}

\maketitle

In the field of strongly correlated electron systems materials with cross-correlated magnetic and
electric properties, called magnetoelectrics, are intensely discussed because of their potential
for controlling magnetic properties by an electric voltage. Among them, the magnetoelectric
multiferroics may be most prominent because due to a coexistence of magnetic and electric order
they can develop particularly pronounced magnetoelectric
interactions.\cite{Eerenstein06,Cheong07a} Recently it was demonstrated that intrinsically strong
(``giant'') magnetoelectric effects are present in the so-called joint-order-parameter
multiferroics in which magnetic long-range order breaks the inversion symmetry and induces a
spontaneous polarization.\cite{Newnham78a,Kimura03a,Hur04a} For example, in TbMnO$_3$,
Ni$_3$V$_2$O$_8$, and MnWO$_4$, a spiral arrangement of spins violates the inversion symmetry and
causes a spontaneous polarization
\begin{equation}\label{eq:spincurrent}
 \mathbf P \propto \mathbf e_{ij}\times(\mathbf S_i \times \mathbf S_j) \; ,
\end{equation}
with $\mathbf e_{ij}$ as unit vector connecting neighboring spins at sites $i$ and $j$ and
$(\mathbf S_i \times \mathbf S_j)$ as vector chirality.\cite{Kenzelmann05a,Lawes05a,Taniguchi06}
The magnitude and direction of $\mathbf P$ are determined by the magnetic order only, so that a
unique correlation between the magnetic and ferroelectric (FEL) order parameters is obtained.
Although the spontaneous polarization is usually small, its robustness\cite{Kimura08a} renders
joint-order-parameter multiferroics interesting for future applications.

An essential feature of any ferroic material is the presence of domains. They determine the
switching of information bits in memory devices and the technological performance of permanent
magnets. At its root, any magnetoelectric interaction in a multiferroic corresponds to an
interaction of its magnetic and electric domains. Hence, understanding giant magnetoelectric
effects means understanding the nature and interactions of the multidomain state in the
joint-order-parameter multiferroics.\cite{Loidl08} However, although domains are known to be
present in these compounds by a variety of (mostly indirect)
experiments\cite{Choi09a,Bodenthin08a,Meier09a,Kimura08a,Kundys08a,Radaelli08a,Cabrera09a} the
prime target of previous investigations was actually to {\it remove} these domains by converting
the samples to or in between single-domain states.

Therefore, none of the existing publications addresses the three-dimensional distribution of
domains in a joint-order-parameter multiferroic and the effects guiding their formation. With
this, essential questions regarding the nature of the domains remain unclear. For example, to what
extent can a domain actually be called ferroelectric if the spontaneous polarization is
magnetically induced?

In this report, the three-dimensional topology of domains in the joint-order-parameter
multiferroic MnWO$_4$ is resolved by optical second harmonic generation (SHG). The spatial
distribution of the domains, their response to external fields, and annealing procedures reveal
some features that are uniquely associated to a magnetic domain state and others that point
unambiguously to ferroelectric domains. The concept of ``multiferroic hybrid domains'' is thus
introduced whereas a description in terms of ``magnetic domains'' or ``electric domains'' or even
``magnetic domains coexisting with electric domains'' is no longer appropriate.

The evolution of the multiferroic phases and their domains in joint-order-parameter compounds can
be described by Landau theory. In the case of spin-spiral systems it was shown that a coexistence
of two magnetic order parameters is required for obtaining
multiferroicity.\cite{Toledano09a,Toledano09b,Harris07a,Lawes08a} The first magnetic order
parameter generally guides the system into a phase with an incommensurate non-polar spin
arrangement. Upon cooling another magnetic transition described by a second magnetic order
parameter evolves, so that the combination of both breaks the inversion symmetry of the crystal.
As a consequence, a spontaneous polarization according to Eq.~(\ref{eq:spincurrent}) can emerge.

For the investigation of the local structure of domains in spin-spiral ferroelectrics the choice
of MnWO$_4$ as a model system suggests itself. On the one hand, its magnetic lattice is rather
simple, because no rare-earth and just one kind of transition-metal ion contributes to the
magnetic order. On the other hand, the rich magnetic phase diagram includes transitions from the
multiferroic phase to neighboring ordered or disordered states which allows one to probe memory
effects in either.\cite{Arkenbout06,Taniguchi06}

\begin{figure}
\includegraphics[width=8.7cm,keepaspectratio,clip]{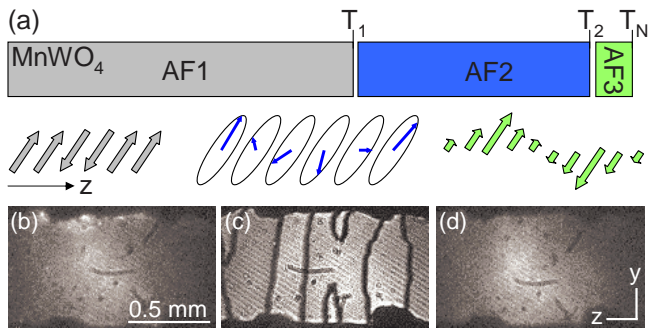}
\caption{\label{fig:1}(a) Magnetic phase transitions in MnWO$_4$. Below the N\'eel temperature
$T_N$ the Mn$^{3+}$ moments display collinear incommensurate long-range order within the easy
plane. At $T_2$ an additional transverse spin component orders which leads to a helical
incommensurate arrangement of spins. In the magnetic ground state below $T_1$ a commensurate
up-up-down-down spin ordering is obtained. The spin spiral in the AF2 phase induces a spontaneous
electric polarization along the $y$ axis while the AF1 and AF3 phases are not multiferroic. (b --
d) Spatially resolved SHG measurements in the AF1 to AF3 phases. A domain structure (with black
lines revealing domain walls) is detected in the multiferroic AF2 phase only. $x$, $y$, and $z$
represent a Cartesian coordinate system, approximating the monoclinic unit cell with lattice
parameters $a=4.83$~\AA, $b=5.76$~\AA, $c=4.99$~\AA \, and $\beta=91.1^\circ\approx 90^\circ$.}
\end{figure}

Three magnetically ordered phases determine the low temperature behavior of MnWO$_4$. In the AF3
phase right below $T_N=13.5$~K incommensurate antiferromagnetic (AFM) ordering of the Mn$^{3+}$
moments described by the order parameter $\mathcal O_{\mathrm{AF3}}$ occurs. The spins align in a
collinear way, pointing along the easy-axis within the $xz$ plane so that they enclose an angle of
about $34^\circ$ with the $x$ axis of the monoclinic crystal as illustrated in
Fig.~\ref{fig:1}(a).
Their magnitude is sinusoidally modulated, leading to a two-dimensional
incommensurate spin-density wave with $\mathbf k =
(-0.214,\frac{1}{2},0.457)$.\cite{lautenschlaeger93a}

In the multiferroic AF2 phase below $T_2=12.7$~K an additional transverse spin component orders
and the spin density wave becomes an elliptical spin spiral while retaining $\mathbf k =
(-0.214,\frac{1}{2},0.457)$ as wave vector. The related magnetic order parameter $\mathcal
O_{\mathrm{AF2}}$ evolves through a second-order phase transition at $T_2$ and coexists with
$\mathcal O_{\mathrm{AF3}}$. As detailed above, simultaneous presence of the two magnetic order
parameters breaks the inversion symmetry and induces a spontaneous polarization, here
\begin{equation}\label{eq:pol}
 P_y\propto\mathcal O_{\mathrm{AF3}}\mathcal O_{\mathrm{AF2}}
\end{equation}
along to the $y$ axis, and establishes multiferroicity in MnWO$_4$. Note that $\mathcal
O_{\mathrm{AF2}}$ can be reoriented below $T_2$ while $\mathcal O_{\mathrm{AF3}}$ is
frozen.\cite{Harris07a,Lawes08a,Toledano09b} Thus, any reversal of $\mathcal O_{\mathrm{AF2}}$ is
coupled one-to-one to a reversal of $P_y$.\cite{Yamasaki06a,Tokunaga09a}

In the AF1 phase, MnWO$_4$ displays collinear commensurate AFM order. The associated wave vector
is $\mathbf k=(\pm\frac{1}{4},\frac{1}{2},\frac{1}{2})$ and describes an up-up-down-down spin
structure with spins aligned along the easy-axis. In agreement with Eq.~(\ref{eq:spincurrent}),
the magnetic order parameter $\mathcal O_{\mathrm{AF2}}$ and, thus, $P_y$, are quenched at the
first-order ${\rm AF2}\to{\rm AF1}$ transition at $T_1=7.6$~K while $\mathcal O_{\mathrm{AF3}}$
remains.

Because of the intricate interplay of the order parameters $\mathcal O_{\mathrm{AF2}}$ and
$\mathcal O_{\mathrm{AF3}}$ the number and types of domains for the three phases will differ. A
powerful technique for investigating domain structures in systems with magnetic or electric order
is second harmonic generation (SHG). Optical SHG describes the induction of a light wave at
frequency $2\omega$ by a light wave at frequency $\omega$. The process picks up the symmetry
changes imposed by the long-range order by coupling directly to the corresponding order parameter
and its different orientation in different domains. A detailed discussion of the technical aspects
of SHG in ferroic systems in general\cite{Fiebig05b} and in MnWO$_4$ in particular\cite{Meier09a}
was already published so that we restrict ourselves here to the application of SHG for imaging the
domain structure of MnWO$_4$.

Obviously, SHG contributions arising at $T_N$ probe the magnetic order parameter $\mathcal
O_{\mathrm{AF3}}$, while those emerging at $T_2$ involve coupling to $\mathcal O_{\mathrm{AF2}}$
and, because of Eq.~(\ref{eq:pol}), to $P_y$. As overview, Figs.~\ref{fig:1}(b)-(d) show spatially
resolved measurements of the SHG intensity in the AF1, AF2, and AF3 phase, respectively. While the
AF1 and the AF3 phases reveal a homogeneous distribution of SHG intensity, the multiferroic AF2
phase exhibits about ten curved black lines on an otherwise homogeneous background. Such lines are
a hallmark for the presence of domains with opposite orientation of the corresponding order
parameter. Because of the linear coupling to the order parameter, the SHG light field experiences
a sign reversal corresponding to a $180^{\circ}$ phase shift when crossing the domain wall. This
leads to destructive interference in the vicinity of the domain wall and, therefore, to the black
lines. According to Fig.~\ref{fig:1} such $180^{\circ}$ domains are present in the multiferroic
AF2 phase only while being absent in the AF1 and AF3 phases.

In order to understand this observation we have to analyze the symmetry of the AF1 to AF3 phases.
The monoclinic paraelectric and paramagnetic phase of MnWO$_4$ belongs to the point-group
$2_y/m_y1'$. This point symmetry is preserved by the commensurate up-up-down-down spin
configuration in the AF1 phase and also by the incommensurate spin-density wave in the AF3 phase
because the magnetic order breaks translation symmetries only.\cite{lautenschlaeger93a} Therefore,
a formation of domains with different orientation of the order parameter should indeed not
occur.\cite{footnote} In contrast, the multiferroic AF2 phase possesses the point symmetry $2_y1'$
with half the number of point-symmetry operations as in the group $2_y/m_y1'$. As observed, this
leads to two domains with an opposite sign of the magnetic order parameter $\mathcal
O_{\mathrm{AF2}}$ corresponding to an opposite magnetic vector chirality and, via
Eq.~(\ref{eq:pol}), to an opposite sign of the spontaneous polarization $P_y$.

\begin{figure}
\includegraphics[width=8.7cm,keepaspectratio,clip]{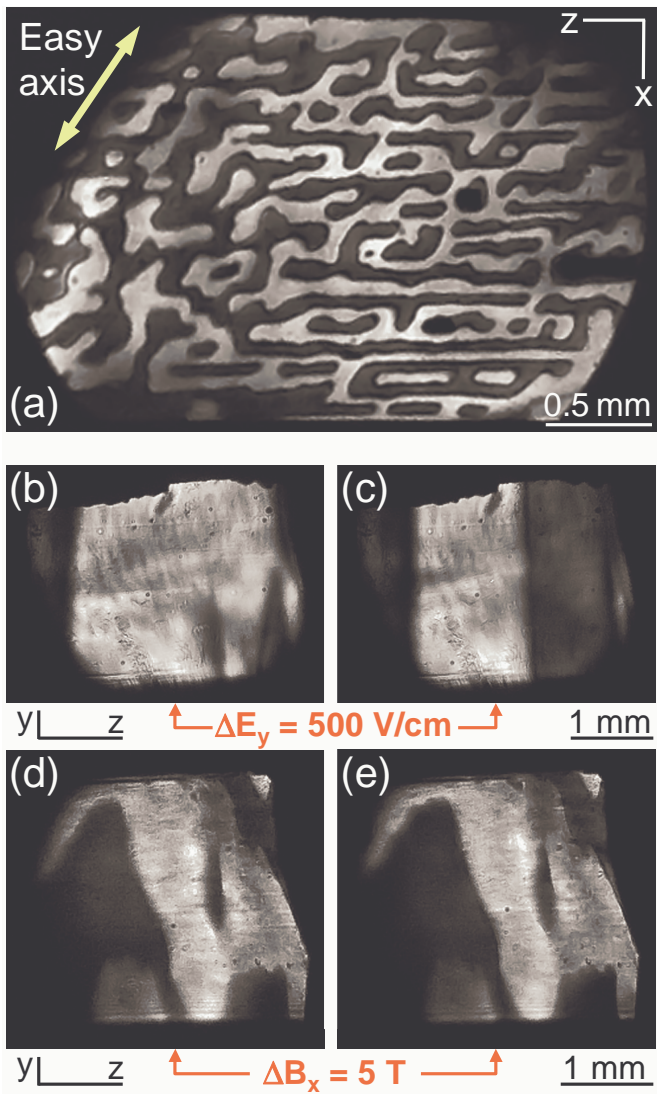}
\caption{\label{fig:2} Spatially resolved SHG images reveal the hybrid nature of the multiferroic
domains in MnWO$_4$. (a) Domain structure in the $xz$ plane. A pronounced elongation along the
magnetic easy axis of the crystal (yellow arrow) and the bubble topology reflect the magnetic
aspect of the domains. (b, c) Domain structure in electric fields of (b) 167~kV/cm and (c)
217~kV/cm applied along the $y$ axis. (d, e) Domain structure in magnetic fields of (d) 0~T and
(e) 5~T applied along the $x$ axis. Only the electric field affects the distribution of the
domains, thus reflecting their electric aspect.}
\end{figure}

In the following, we will focus on the discussion of the $180^{\circ}$ domains in the multiferroic
AF2 phase. First of all, we will investigate the topology of the AF2 domains and their response to
external poling fields. Figure~\ref{fig:2}(a) shows the distribution of the AF2 domains in the
$xz$ plane after zero-field cooling.
We find a labyrinth-like arrangement of dark and bright
areas, being distributed with approximately equal proportions across the sample. The different
level of brightness corresponds to opposite magnetic vector chirality and spontaneous polarization
according to Eq.~(\ref{eq:spincurrent}). In contrast to Fig.~\ref{fig:1}(c) a different brightness
between opposite domains (in addition to the mere domain walls) is observed because of the
interference of the SHG light field arising from the multiferroic order with a homogeneous SHG
light field generated by the crystal lattice.\cite{Fiebig05b} The topology of the domains in
Fig.~\ref{fig:1}(a) reveals two preferential directions: along the $z$ axis and along a line
including an angle of about $34^{\circ}$ with the $x$ axis.

Figure~\ref{fig:2}(a) clearly expresses the magnetic origin of the domain structure. First, the
texture of the domains strikingly resembles the patterns of bubble and stripe domains, universally
attributed to modulated phases with a preselected equilibrium periodicity.\cite{Seul95a} In
MnWO$_4$ it is the long-range magnetic interaction that stabilizes the modulated AF2 phase and the
periodicity is determined by the magnetic wave vector $\mathbf k$. Second, the preferential
direction of the domain walls parallel to the arrow in Fig.~\ref{fig:2}(a) is in striking
coincidence with the direction of the magnetic easy-axis of MnWO$_4$ which encloses an angle of
$34^{\circ}-37^{\circ}$ with the $x$ axis in the $xz$ plane.\cite{lautenschlaeger93a}

In Figs.~\ref{fig:2}(b -- e) the response of the domain structure in the $yz$ plane to electric
and magnetic fields is shown. It is obvious that with an electric field applied along the $y$ axis
small changes in the order of 1~kV/cm lead to pronounced changes in the topology of the AF2
domains. In contrast, a magnetic field of up to 5~T applied along the $x$ axis does not alter the
domain structure at all [Fig.~\ref{fig:2}(d) and \ref{fig:2}(e)], and the same holds for fields
applied along the $y$ or $z$ axis. Figures~\ref{fig:2}(b -- e) thus emphasize the electric nature
of the AF2 domain structure.

We therefore conclude that the AF2 domains exhibits hallmarks of both a magnetically {\it and} an
electrically ordered state. The {\it topology} is determined by the \textit{magnetic} character of
the domains, whereas the \textit{field response} reflects the \textit{electric} character. Because
of this bilateral nature, only a denomination of the AF2 domains as {\it multiferroic hybrid
domains} seems appropriate whereas a description in terms of ferroelectric or antiferromagnetic
remains incomplete. This is in stark contrast to earlier work\cite{Meier09a} where magnetic and
ferroelectric domains where still considered as strictly separate entities. It was already pointed
out that the improper nature of the ferroelectric polarization implies rigid coupling to the
magnetic order parameters.\cite{Yamasaki06a,Tokunaga09a} However, this seems to be the first work
where the consequences of this relation on the level of the domain structure is revealed. Only
this leads to the realization that an extended concept of multiferroic hybrid domains is required.

\begin{figure}
\includegraphics[width=8.7cm,keepaspectratio,clip]{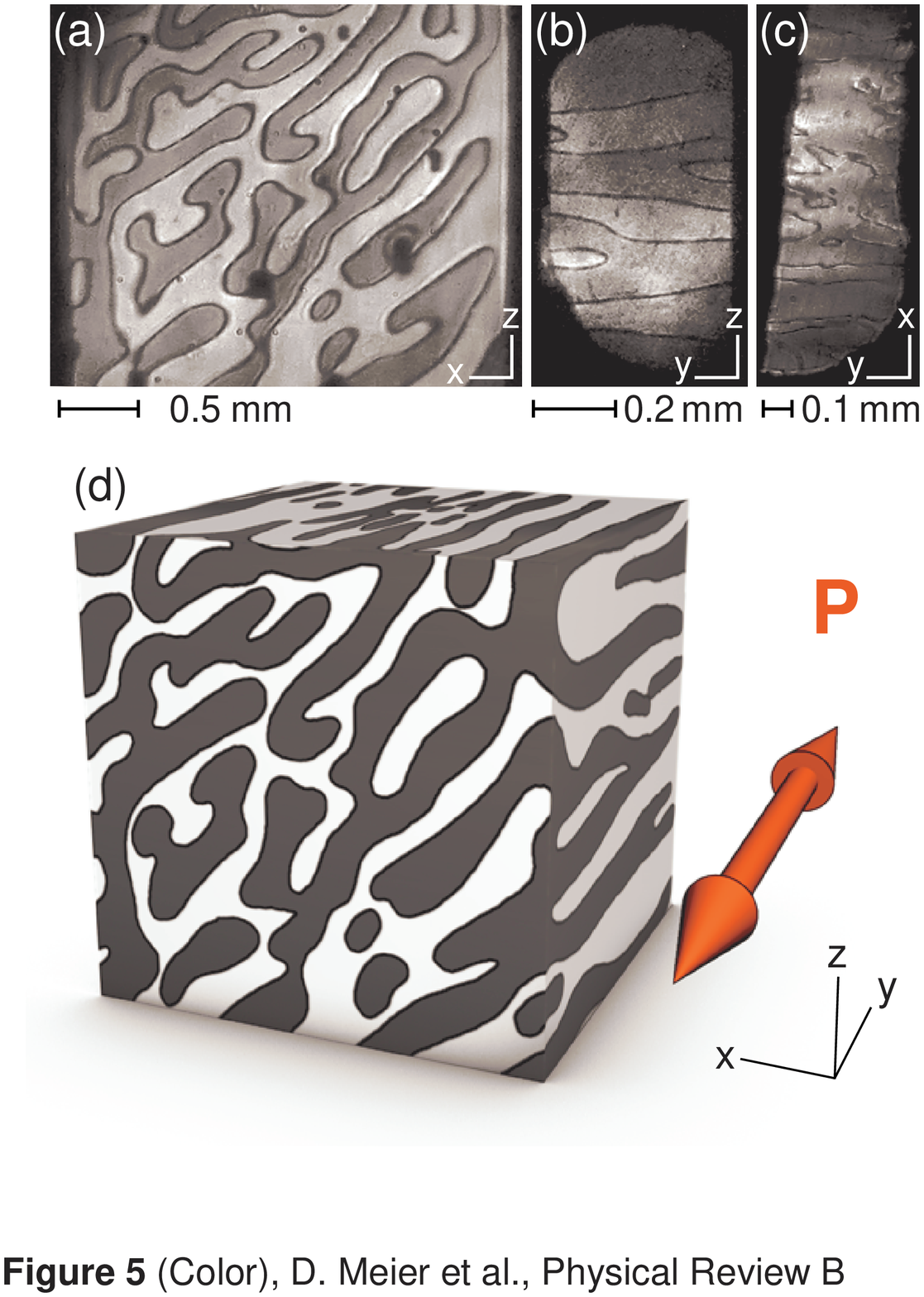}
\caption{\label{fig:3} Three-dimensional distribution of the multiferroic domains. (a -- c)
Domains in the $xz$, $yz$, and $xy$ plane of MnWO$_4$ samples taken from the same batch. (d)
Three-dimensional visualization of the multiferroic domain structure in (a -- c).}
\end{figure}

The full three-dimensional topology of the multiferroic domains is presented in Fig.~\ref{fig:3}
for three differently oriented MnWO$_4$ samples obtained from the same batch.
Figure~\ref{fig:3}(a) shows the domain structure in the $xz$ plane. Like in Fig.~\ref{fig:2}(a),
dark and bright regions correspond to the two possible multiferroic domains with opposite order
parameters. We can see the aforementioned elongation of domains along the magnetic easy-axis with
the spontaneous polarization $P_y$ pointing into and out of the plane. The propagation of domain
walls along the direction of the spontaneous polarization is revealed by Figs.~\ref{fig:3}(b) and
~\ref{fig:3}(c). Interestingly, the domain walls, here again indicated by the black lines,
continue rather straight along the $y$ axis of the crystal. The domains tend to form platelets in
planes defined by the magnetic easy-axis and the direction of the spontaneous polarization. With a
lateral extension evaluated as $170 \pm 60$~$\mu$m\cite{hubert98} the domains are surprisingly
large. For an improved visualization of the anisotropic multiferroic domain structure,
Fig.~\ref{fig:3}(d) shows a three-dimensional simulation based on the distribution of domains in
Figs.~\ref{fig:3}(a -- c). The respective domain structure were projected onto three faces of a
cuboid and mended minimally at the edges.

\begin{figure}
\includegraphics[width=8.7cm,keepaspectratio,clip]{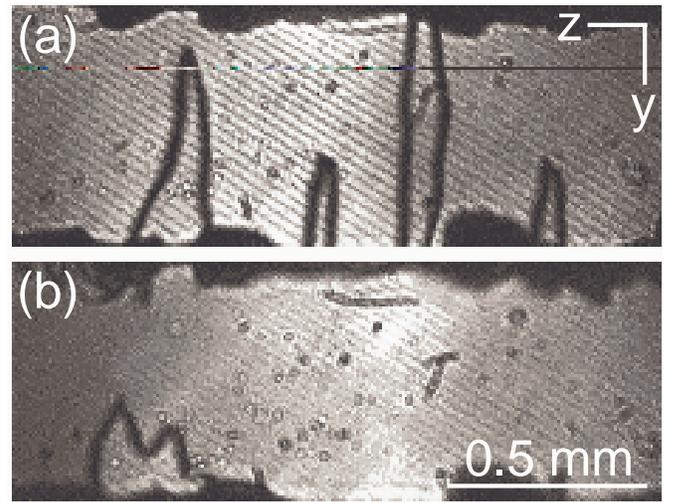}
\caption{\label{fig:4} Response of the multiferroic domains to a temperature annealing cycle
through the paramagnetic phase. (a) Domain structure after initial zero-field cooling from room
temperature to the multiferroic AF2 phase. (b) Domain structure after subsequent application of
the annealing cycle.}
\end{figure}

In the following we consider the dynamic aspects in the formation of the domains by applying
annealing procedures across the boundaries limiting the multiferroic phase. First of all, we
consider the effect of temperature-annealing in an order $\leftrightarrow$ disorder cycle from the
AF2 phase to the paramagnetic state above $T_N$ and back to the multiferroic phase. Initially, the
sample was zero-field-cooled from room temperature into the AF2 phase which leads to the SHG image
shown in Fig.~\ref{fig:4}(a).
In agreement with Fig.~\ref{fig:1}, a multiplicity of domains is
obtained. Subsequently, the order $\leftrightarrow$ disorder annealing cycle was applied which
leads to the SHG image shown in Fig.~\ref{fig:4}(b). Most of the domain walls have vanished and
only a small fraction of the MnWO$_4$ remains in a domain state opposite to the rest of the
sample. Apparently, the annealing procedure tends to drive the sample towards a single domain
state which was confirmed by repeating the experiment more than ten times.

Figure~\ref{fig:4} leads to two conclusions. First, no memory effect is observed in the order
$\leftrightarrow$ disorder annealing cycle. Second, while conventional ferroelectrics tend to form
a multiplicity of domains for minimizing electric stray fields, an ideal antiferromagnets tends to
approach a single-domain state.\cite{Li56a} Hence, just like the domain structure itself
(Fig.~\ref{fig:2}) the dynamic domain topology is dominated by the magnetic aspect of the hybrid
multiferroic order.

In the second step, we investigated the effect of an order $\leftrightarrow$ order annealing cycle
from the AF2 phase to the AF1 phase and back  to the multiferroic state. This transition is of
particular interest because of a FEL memory effect reported
earlier:\cite{Arkenbout06,Taniguchi09a} A single-domain state in the multiferroic phase is
memorized in the non-polar AF1 phase and reemerges when reentering the AF2 phase. However, all
current data were gained by integral techniques such as pyroelectric current measurements, so that
it is not known to what extend the memory effect applies to the domain structure of the
multiferroic multidomain state.

\begin{figure}
\includegraphics[width=8.7cm,keepaspectratio,clip]{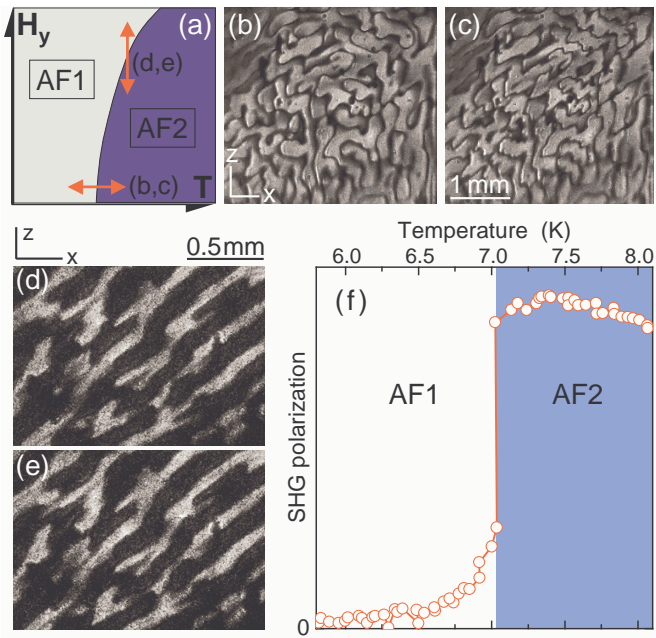}
\caption{\label{fig:5} Response of the multiferroic domains to magnetic-field ($H$) and
temperature ($T$) annealing cycles through the AF1 phase. (a) Sketch of the $(H, T)$ phase diagram
with arrows indicating the respective annealing procedure. (b, d) Domain structure after initial
zero-field cooling from room temperature to the multiferroic AF2 phase. (c, e) Domain structure
after application of the respective annealing cycle. (f) Temperature dependence of the SHG signal
($\chi_{yxx}$ component at 1.95~eV)\protect\cite{Meier09a} in a temperature decreasing run.
Nonzero SHG yield in the AF1 phase points to a residual polar contribution in the AF1 phase as
probable basis of the topological memory effect revealed in panels (b -- e).}
\end{figure}

This is investigated in Fig.~\ref{fig:5} which compares the domain structure of the AF2 phase
before [panels (b), (d)] and after [panels (c), (e)] the annealing cycle through the AF1 phase.
Figure~\ref{fig:5}(a) illustrates that the phase boundary between the AF2 and the AF1 phase can be
crossed by magnetic-field or temperature tuning. Figures~\ref{fig:5}(b) and ~\ref{fig:5}(c) reveal
the effects of thermal annealing. Minor changes in the roughness of the domain walls and, for a
small fraction of the walls, shifts in the order of 100~$\mu$m are observed but the general
distribution of the domains is preserved. Annealing in a magnetic field along the $y$ axis reveals
no changes at all in the domain structure. Hence, not only a single-domain state is memorized in
the non-chiral non-polar AF1 phase --- actually the entire topology of a multidomain state is
preserved. The memory effect is thus much more rigid than established up to now.

Figure~\ref{fig:5} immediately raises the question for the origin of such a memory effect. Pinning
of the domain structure by structural defects can be excluded. This mechanism would also preserve
the domain structure in the order $\leftrightarrow$ disorder transition to the paramagnetic phase
contrary to what Fig.~\ref{fig:4} shows. For revealing its origin we measured the temperature
dependence of the magnetic order parameter $\mathcal O_{AF2}$ across the AF2 $\rightarrow$ AF1
transition. As shown in Fig~\ref{fig:5}(f), the SHG contribution that is related to the magnetic
order parameter $\mathcal O_{\mathrm{AF2}}$ does not vanish abruptly at the phase boundary to the
AF1 state. Instead, it diminishes gradually with temperature after an initial step-like decrease.
The remanent SHG signal reveals a coexistence of the AF1 and AF2 phase within a broad temperature
region, enabled by the first-order nature of the transition. Hence, residual nuclei of the AF2
phase explain the ``polar memory'' of MnWO$_4$.\cite{Taniguchi09a} The distribution of the nuclei
must be homogeneous and dense in order to explain pinning of the entire topology of a multidomain
state. Note that~\ref{fig:5}(f) is the first measurement directly proving the existence of polar
inclusions within the AF1 phase, which is possible because of the extraordinary sensitivity of SHG
to nanoscopic inclusions.\cite{Kordel09a}

In summary, we revealed that the magnetically induced ferroelectric phase of the
joint-order-parameter multiferroic MnWO$_4$ forms novel types of domains for which characteristic
features of magnetic and ferroelectric domains are inseparably entangled. Here the denomination as
{\it multiferroic hybrid domain} is introduced. In MnWO$_4$ the {\it topology} of these chimera
domains is determined by their {\it magnetic} nature, whereas the {\it field response} reflects
the {\it electric} character. The three-dimensional distribution of the domains was investigated
and annealing cycles revealed a topological memory effect allowing one to reconstruct the entire
multidomain structure subsequent to quenching it.

The present work should provide a basis for the development of a model explaining the topology of
domains in compounds with magnetically induced ferroelectricity. Regarding long-term application,
a vastly increasing variety of joint-order-parameter multiferroics allowing ferroelectricity
according to Eq.~(\ref{eq:pol}) are at our disposal for multiferroic domain control. The discovery
of mechanisms promoting magnetically induced ferroelectricity different from Eq.~(\ref{eq:pol}) is
only a question of time and may eventually lead us towards high-temperature applications. Because
of the bilateral nature of the multiferroic hybrid domains systems with conical spin spirals allow
one to exert rigid electric-field control of a macroscopic magnetization.\cite{Yamasaki06a} Since
magnetization and polarization are manifestations of the same multiferroic hybrid domain state the
electric field will always act on both. Rapid magnetization reversal with ultrashort
electric-field pulses may thus become feasible.

\begin{acknowledgments}
We thank Steffen Brosseit and BrossBoss Entertainment for designing the 3D domain structure. This
work was supported by the DFG through the SFB608.
\end{acknowledgments}


\begin{thebibliography}{30}
\bibitem{footnote}
Antiphase domains, also called translation domains, may occur.\protect\cite{Meier09a} However,
they result from the violation of only the translation symmetry and do not differ in the
orientation of the order parameter. Therefore, they are not considered henceforth.
\end{thebibliography}
\end{document}